\documentclass{jpp}

\newcommand{\pdff}[2]{\frac{\partial #1}{\partial #2}}

\newcommand{\scr}[1]{_{\mbox{\protect\scriptsize #1}} }

\newcommand{\perme}{\mu_0}

\newcommand{\degrees}{^{\circ}}

\newcommand{\valf}{v_A}

\newcommand{\mcube}{m\textsuperscript{-3}}

\newcommand{\lb}{LaB$_6$~}

\newcommand{\mus}{{\textmu}s}

\newcommand{\bdot}{\dot{B}}
\newcommand{\bz}{B_Z}

\newcommand{\bph}{B_\phi}
\newcommand{\jph}{J_\phi}
\newcommand{\jz}{J_Z}
\newcommand{\erec}{E\scr{rec}}
\newcommand{\eph}{E_\phi}

\newcommand{\bh}{B_H}

\newcommand{\vah}{v\scr{Ah}}
\newcommand{\bred}{B\scr{red}}

\newcommand{\hgas}{\text{H}_2}
\newcommand{\vin}{v\scr{in}}
\newcommand{\vout}{v\scr{out}}
\newcommand{\ms}{M\scr{MS}}
\newcommand{\brec}{B\scr{rec}}
\newcommand{\vdr}{V\scr{drive}}

\newcommand{\vlay}{v\scr{layer}}
\newcommand{\vpile}{v\scr{pile}}
\newcommand{\Vf}{\Phi_f}
\newcommand{\Vs}{\Phi_s}

\usepackage[utf8]{inputenc}
\usepackage[T1]{fontenc}

\usepackage{graphicx}
\usepackage{amsmath}
\usepackage{amssymb}
\usepackage{times}
\usepackage{bm}
\usepackage[square]{natbib}

\usepackage{multirow}
\usepackage{hhline}
\usepackage[table,dvipsnames]{xcolor}
\usepackage{rotating}
\usepackage{array}
\usepackage{textcomp}
\usepackage{float}
\usepackage{upgreek}
\usepackage[italicdiff]{physics}

\definecolor{blue}{rgb}{0,0,1}
\usepackage[colorlinks=true,pdfborder={0 0 0},allcolors=blue]{hyperref}

\graphicspath{{figs/}}

\shorttitle{Shocked Flux Pileup and Magnetic Reconnection}
\shortauthor{J. Olson, J. Egedal, and others}

\title{Regulation of the Normalized Rate of Driven Magnetic Reconnection through Shocked Flux Pileup}

\author{Joseph~Olson\aff{1}\corresp{\email{joseph.olson@wisc.edu}},
    Jan~Egedal\aff{1}, Michael~Clark\aff{1}, Douglass~A.~Endrizzi\aff{1}, Samuel~Greess\aff{1}, 
    Alexander~Millet-Ayala\aff{1}, Rachel~Myers\aff{1}, Ethan~E.~Peterson\aff{1}\aff{2},
    John~Wallace\aff{1}, Cary~B.~Forest\aff{1}
    }

\affiliation{
\aff{1}Department of Physics, University of Wisconsin--Madison, Madison, WI 53706, USA
\aff{2}Plasma Science and Fusion Center, MIT, Cambridge, MA 02139, USA
}

\begin{document}

\maketitle

\begin{abstract}
    Magnetic reconnection is explored on the Terrestrial Reconnection Experiment (TREX) for asymmetric inflow conditions and in a configuration where the absolute rate of reconnection is set by an external drive. Magnetic pileup enhances the upstream magnetic field of the high density inflow, leading to an increased upstream Alfv\'en speed and helping to lower the normalized reconnection rate to values expected from theoretical consideration. In addition, a shock interface between the far upstream supersonic  plasma inflow and the region of magnetic flux pileup is observed, important to the overall force balance of the system, hereby demonstrating the role of shock formation for configurations including a supersonically driven inflow. Despite the specialised geometry where a strong reconnection drive is applied from only one side of the reconnection layer, previous numerical and theoretical results remain robust and are shown to accurately predict the normalized rate of reconnection for the range of system sizes considered. This experimental rate of reconnection is dependent on system size, reaching values as high as 0.8 at the smallest normalized system size applied.
\end{abstract}

\section{Introduction} \label{sec: intro}

Magnetic reconnection is a fundamental process in plasma systems which allows for the magnetic topology to change rapidly, converting stored magnetic energy into plasma energy \citep{Zweibel2016}. While reconnection occurs in a localized diffusion region \citep{Burch2016a}, it often leads to dramatic changes in the macroscopic behavior of a variety of systems including solar flares \citep{Masuda1994}, the Earth's magnetosphere \citep{Phan2000}, and magnetic fusion experiments \citep{Wesson1986}.

In reconnection models the upstream magnetic field $\brec$ is important because its tension sets the acceleration of the reconnection exhaust, yielding an outflow speed $\vout\simeq \valf$, where $\valf=\brec/\sqrt{\mu_0 n_i m_i}$ is the Alfv\'en speed \citep{parker1957}. Furthermore, in fast reconnection not only the outflow speed but also the inflow speed (in the frame of the reconneciton layer) is Alfv\'enic \citep{Birn2001}. The normalized reconnection rate $\alpha=\vin/\vout$ has been studied through numerical simulations with typical values of $\mathcal{O}(0.1)$ \citep{Liu2017} but has also been shown to be weakly dependent on system size \citep{Karimabadi2011,Stanier2015a,Ng2015,Phan2018,SharmaPyakurel2019}. By Faraday's law, the absolute rate is characterized by the inductive electric field such that $\erec=\vin\brec$ and is typically influenced by conditions external to the reconnection region \citep{Axford1969,Axford1984}. Nevertheless, it is expected from theory that the reconnection dynamics regulate the current in the reconnection layer such that $\alpha$ remains fixed \citep{Shay2001}. 

The dynamical interplay between shocks and reconnection has many applications to both astrophysical and space plasmas but has only recently been studied in more depth \citep{Matsumoto2015,Karimabadi2014}. For some driven reconnection scenarios, prevalent when stellar winds interact with planetary magnetospheres, a process called magnetic flux pileup regulates the upstream magnetic field $\brec$ such that the ratio of the forced inflow speed and the outflow speed is consistent with the normalized rate \citep{Dorelli2003a}. Examples include the transition from the supersonic solar wind which is compressed at the Earth's bow shock to the magnetosheath upstream of reconnection sites in the dayside magnetopause \citep{Moretto2005,Oieroset2019,Dorelli2019}.

Despite its importance, experimental studies of flux pileup and the role of shock formation are still limited. At dominant plasma pressure, $\beta= nT/(B^2/2\mu_0)\gg1$, flux pileup  has been inferred during the collision of laser produced plasma bubbles \citep{Fiksel2014} in qualitative agreement with numerical models \citep{Fox2011}, as well as in Z-pinch experiments driven by exploding wire arrays \citep{Suttle2016}. For both scenarios, the strong drive yields transient fast reconnection at relatively high Lundquist numbers $S= \mu_0 L \valf/\eta\simeq 10^3$, where $\eta$ is the electrical resistivity and $L$ is the system size. Meanwhile, for $\beta<1$ pileup has only been demonstrated during slower Sweet-Parker-like reconnection \citep{parker1957} between coalescing flux ropes with $S\simeq30$ \citep{Intrator2009}. For all cases above, no shock formation was reported though it likely plays a role in the observed flux pileup. 

In this paper, we present a quantitative experimental study of magnetic flux pileup and the normalized reconnection rate. The study was performed on the Terrestrial Reconnection Experiment (TREX), which operates at the Wisconsin Plasma Physics Laboratory (WiPPL) \citep{Forest2015}. The drive is applied from only one inflow, causing the reconnection layer to move super-Alfv\'enically into the opposing inflow, providing a unique setup that permits the study of magnetic pileup and the associated shock formation under low-collisional conditions, $S\sim10^3$--$10^4$, and where the plasma dynamics are dominated by the magnetic field pressure, $\beta\simeq 0.1$. Furthermore, the large reconnection layer size (half-length of $L\simeq 0.8$ m) facilitates direct probing by simple electrostatic and magnetic diagnostics. Super-Alfv\'enic plasma flows and shock formation are common in heliospheric and astrophysical settings, and the TREX configuration enables a detailed experimental demonstration of the role of shocked flux pileup in regulating the normalized rate during supersonically driven magnetic reconnection.

\section{Experimental Configuration} \label{sec: config}

\begin{figure}
	\centering
    \includegraphics[width=1.0\textwidth]{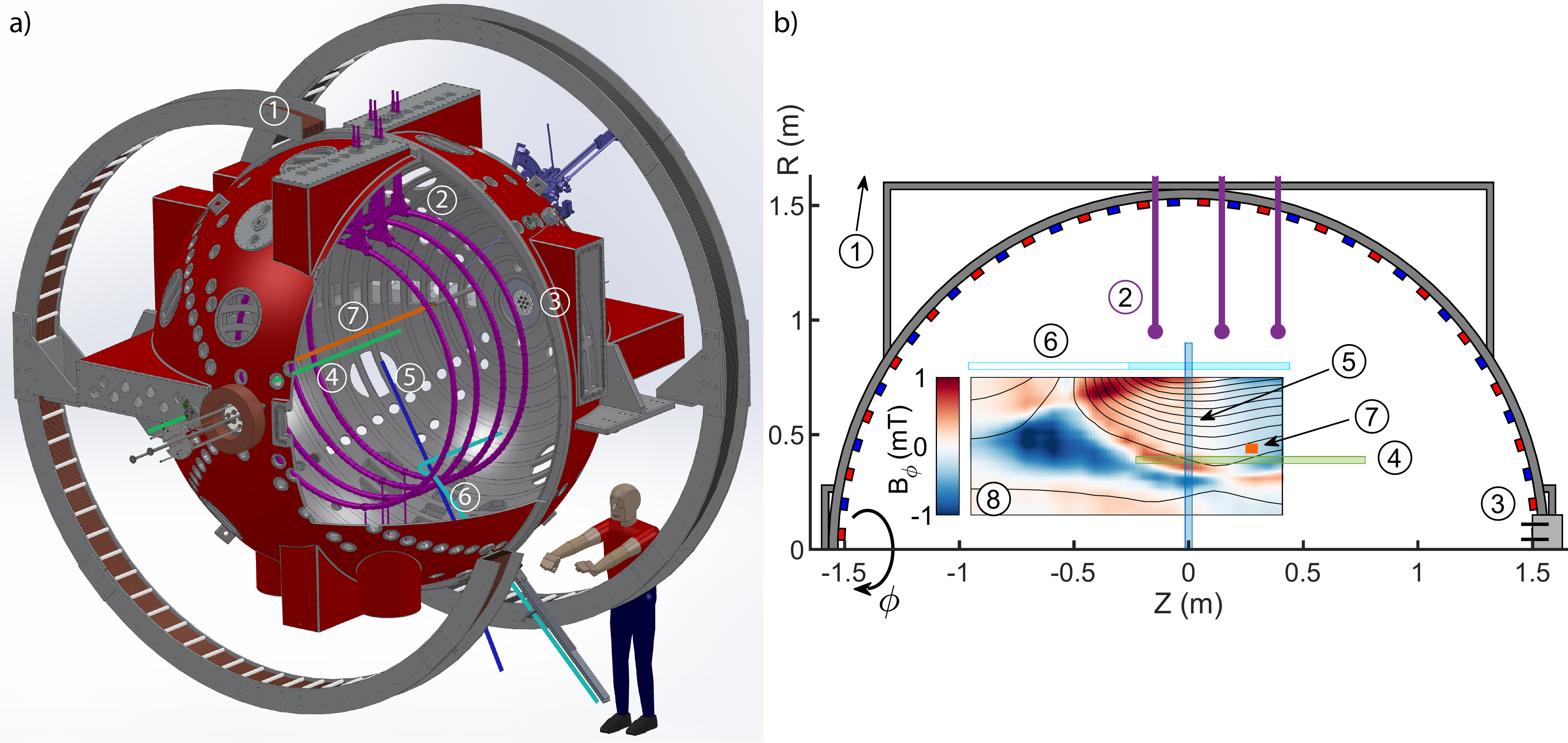}
	\caption{\textit{(a)} A 3D CAD rendering and \textit{(b)} poloidal cross section of the TREX configuration. The primary hardware consists of \textit{(1)} the external Helmholtz coil (at $R=2$~m), \textit{(2)} the internal reconnection drive coils, and \textit{(3)} the plasma gun array added to the 3 m spherical vacuum vessel. The diagnostics consist of three linear $\bdot$ arrays, \textit{(4)} the linear probe, \textit{(5)} the speed probe, and \textit{(6)} the hook probe, as well as \textit{(7)} the $T_e$ probe. The colored segments indicate each probe's spatial coverage.	\textit{(8)} An example 2D profile of the toroidal Hall magnetic fields as measured by probe \textit{(6)} (taken from figure~\ref{f: hook}(p).}
	\label{f: schem}
\end{figure}

The experimental setup is shown in figure~\ref{f: schem}, outlining the location of the primary TREX components and diagnostics used in the following analysis. The configuration closely resembles that presented in \citet{Olson2016}, but with a new reconnection drive system consisting of up to four 0.92~m radius reconnection drive coils pulsed by a low-inductance 1~mF capacitor bank. A 10~ms plasma is initiated by an array of pulsed plasma guns \citep{Fiksel1996} with a steady state, axial magnetic field from the Helmholtz coil. For these experiments, three reconnection drive coils, located at $Z=\pm0.15$ and 0.40~m, are pulsed opposite the Helmholtz field. As the current increases and new magnetic flux is injected by the coils, a reconnection current layer forms and is driven towards the central axis. Figure~\ref{f: schem}(8) shows a 2D profile of the resulting toroidal Hall magnetic fields and poloidal field lines for the configuration presented in figure~\ref{f: rec}. The stronger drive permits increased voltages at the coils by a factor of 10--100 compared to the earlier TREX experiments. In addition, the background plasma temperature and density is naturally lower from the plasma guns ($T_e\sim4$~eV and $n_e\lesssim10^{18}$~\mcube) compared to the previously used \lb cathodes ($T_e\sim10$~eV and $n_e\gtrsim10^{18}$~\mcube).

The experiments utilize a combination of magnetic and electrostatic diagnostics. The colored segments in figure~\ref{f: schem}(b) show the spatial extent for each probe assembly in the poloidal $R\text{--}\phi$ plane. There are two linear arrays consisting of 14 individual 3-axis $\bdot$ probes, the linear probe and the hook probe. The linear probe, spanning 1~m in the $Z$ direction at $R=0.4$~m, can be scanned to different $Z$ locations. The hook probe enters the device at $Z=-0.25$~m parallel to $R$ before bending 90$\degrees$ to cover 0.8~m in the $Z$ direction at varying $R$ positions. This array can reach from $Z=-0.95$--$0.45$~m as indicated by the light blue segments in figure~\ref{f: schem}(b) by rotating 180$\degrees$ around its shaft. Additionally, the speed probe is a stationary array of single-axis $\bdot$ probes spanning from $R=0$--$0.9$~m at $Z=0$~m while measuring $\partial B_Z/\partial t$. Finally, located at $R=0.4$~m and scanned in $Z$, a similar electrostatic probe to one used in previous TREX work measures the full $I$-$V$ plasma characteristic by individually biasing 16 closely spaced Langmuir electrodes. This $T_e$ probe and the hook probe are toroidally offset from the linear probe by 18$\degrees$ and 145$\degrees$ respectively.

\section{Reconnection Geometry} \label{sec: rec geo}

\begin{figure}
	\centering
	\includegraphics[width=\textwidth]{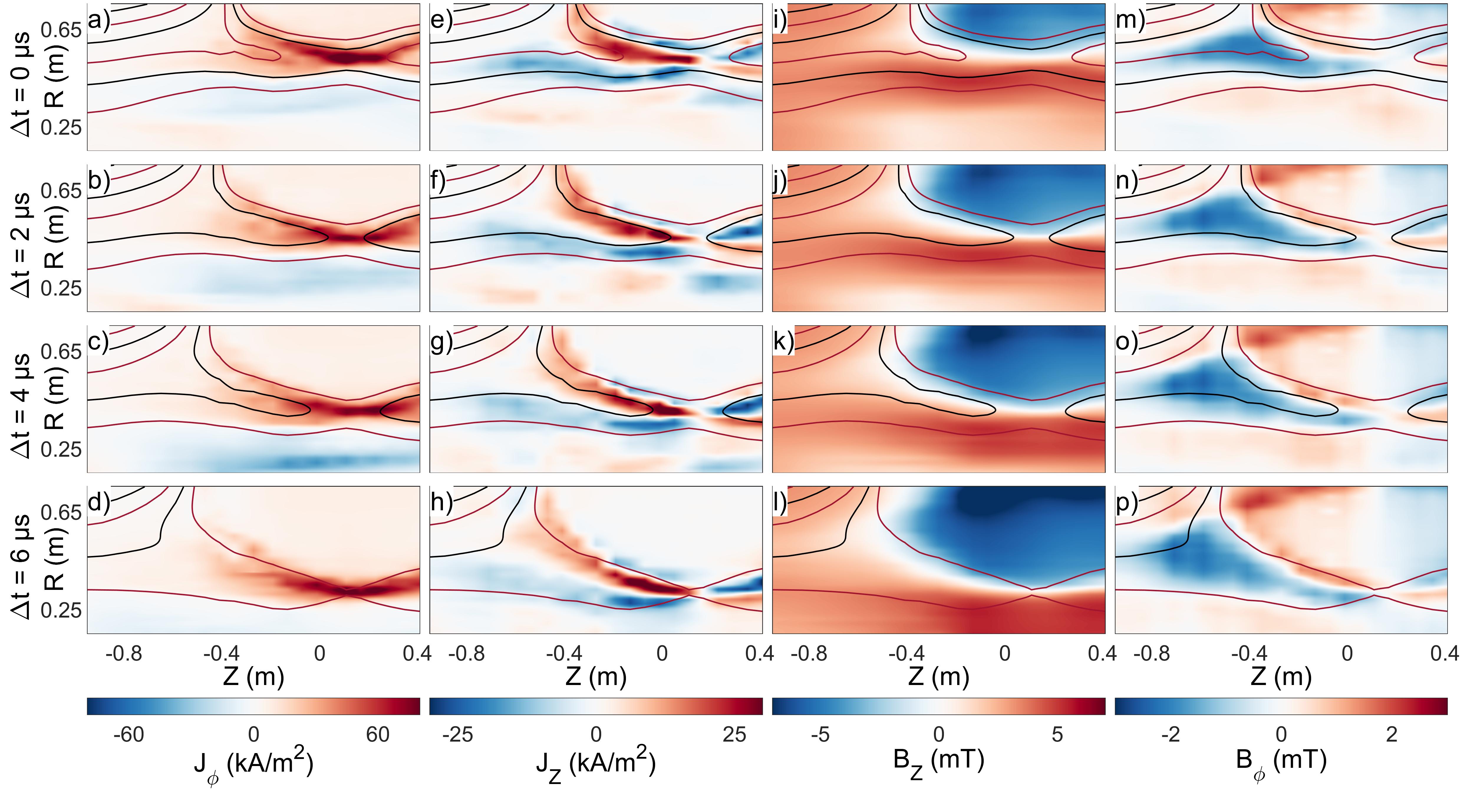}
	\caption[Translation of the reconnection geometry]{Profiles of (a-d) the toroidal current density $\jph$, (e-h) the poloidal current density $\jz$, (i-l) the reconnecting magnetic field $\bz$, and (m-p) the toroidal Hall magnetic field $\bph$ at different times during a shot in $\hgas$ with $\bh=5$~mT and $\vdr=5$~kV. The current layer geometry remains roughly constant while propagating from $R=0.55$~m to $R=0.31$~m.}
	\label{f: hook}
\end{figure}

The reconnection dynamics driven in the TREX configuration are highly reproducible for a given configuration. Figure~\ref{f: hook} shows example profiles of the reconnection geometry for a discharge in Hydrogen gas ($\hgas$) with a drive voltage $\vdr=5$~kV and Helmholtz field $\bh=5$~mT. The evolution of the magnetic configuration is characterized by scanning the hook probe to different radial locations over $\sim$50 discharges with the same experimental settings. 
At $\Delta t=0$~{\mus}, the reconnection geometry is already established. As time increases from top to bottom, the reconnection layer of high toroidal current density $\jph$ propagates from $R=0.55$~m to $R=0.31$~m with velocity $\vlay\simeq40$~km/s. This current layer separates the two reconnection inflows with reversed axial magnetic field $\bz$ shown in figure~\ref{f: hook}(i-l). At $\Delta t=4$~{\mus} in figure~\ref{f: hook}(c), a blue ribbon of negative $\jph$ current is observed at $R\simeq 0.23$~m corresponding to the shock formation and magnetic pileup described in \S\ref{sec: erec} and \S\ref{sec: shock}. Meanwhile, the in-plane current $\jz$ gives rise to the toroidal Hall magnetic field $\bph$.

The overlaid field lines are contours of constant magnetic flux, coinciding with the poloidal magnetic field lines. All panels consider the same set of contour levels such that the motion of the field lines can be followed in time. Within the two inflow regions, the field lines are all observed to move downwards, with those above the reconnection layer moving faster than the layer and those below the layer moving slower than the layer such that in the frame of the reconnection layer, the rate at which field lines are reconnected is roughly constant in time. While the length of the reconnection exhaust does increase slightly in time, the structure of the inner part of the reconnection region is steady as it translates to lower values of $R$.

\section{Reconnection Electric Field} \label{sec: erec}

The speed probe [figure~\ref{f: schem}(5)] provides data critical to evaluate the speeds for the shock and reconnection layers. In addition, the analysis of the speed probe data provides detailed information on the temporal evolution of the inductive electric field as the reconnection layer transits the radial cross section.

\begin{figure}
	\centering
	\includegraphics[width=0.7\textwidth]{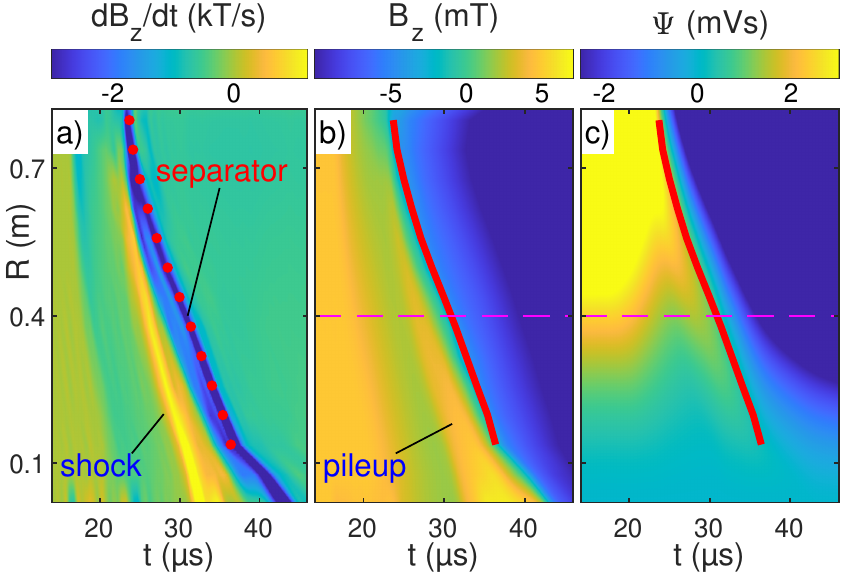}
	\caption[The reconnection and shock layers from the speed probe]{(a) Profile of $\partial\bz/\partial t$ along $R$ versus time from the speed probe. The reconnection current layer (separator) corresponds to the band of negatively peaked $\partial\bz/\partial t$. A shock front, corresponding to the positive peak in the profile, leads the current layer. (b) Profile of $\bz$, integrated from (a). A region of pileup follows the shock front at lower $R$. (c) Profile of the magnetic flux $\Psi$, calculated from (b).}
	\label{f: psi}
\end{figure}

The magnetic pickup loops comprising the speed probe are arranged to measure $\partial B_Z/\partial t$ along a vertical chord at $Z=0$~m. An example of $\partial B_Z/\partial t$ as a function of $(t,R)$ acquired in a single shot is shown in figure~\ref{f: psi}(a). According to Ampere's law we have $\mu_0\jph = \partial B_R/\partial Z - \partial B_Z/\partial R$. Because of the elongated shock and reconnection geometry observed in figure~\ref{f: hook} it is clear that, in general, $|\partial B_Z/\partial R| \gg |\partial B_R/\partial Z|$. It then follows that
\begin{equation}
	\mu_0\jph \simeq - \pdff{B_Z}{R} \simeq -\frac{1}{\vlay} \pdff{B_Z}{t} \, .
\end{equation}
The profile in figure~\ref{f: psi}(a) therefore provides a direct image of the toroidal current as a function of $(t,R)$, where the reconnection separator (current layer) is identified as the negative ridge of  $\partial B_Z/\partial t$ and the shock front leading the separator is the positive ridge of $\partial B_Z/\partial t$. The local slopes of these fronts in the $(t,R)$-plane yield the speeds $\vpile$ and $\vlay$, for the shock and reconnection layer, respectively. 

Taking into account the background Helmholtz magnetic field, $\bh$, the profile of $B_Z$ in figure~\ref{f: psi}(b) is readily obtained as
\begin{equation}
	B_Z(t,R) = \bh + \int_0^t \frac{\partial B_Z}{\partial t'} dt' \, .
\end{equation}
We have marked the region of magnetic pileup corresponding to the magnetic field compression downstream of the shock front. In addition, the red line indicates the trajectory of the separator, $R_X(t)$. For our purposes the magnetic flux function is defined as 
\begin{equation}
	\Psi(t,R) = 2\pi \int_0^R R' B_Z(t,R')dR' \, ,
\end{equation}
which represents the \textit{total} poloidal flux through a toroidal loop of radius $R$. In turn, $\Psi(t,R)$ is readily computed and shown in figure~\ref{f: psi}(c).

\begin{figure}
	\centering
	\includegraphics[width=\textwidth]{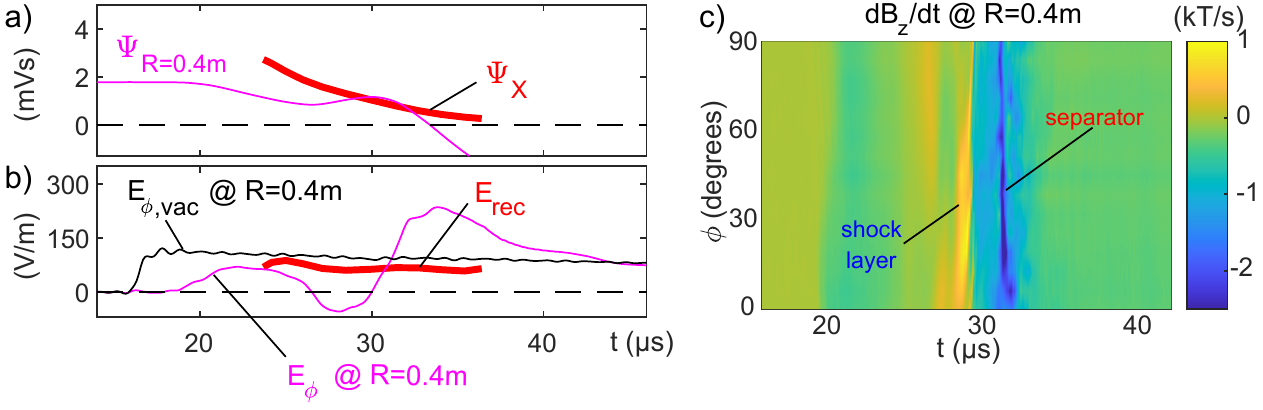}
	\caption[Calculating the reconnection electric field]{(a) Time trace of $\Psi$ at $R=0.4$~m [along the dashed pink line in figure~\ref{f: psi}(c)] and the magnetic flux evaluated along the separator trajectory [red line in figure~\ref{f: psi}(c)]. (b) Inductive electric field $\eph$ evaluated at $R=0.4$~m and along the separator trajectory, denoted as the reconnection electric field $\erec$. $\eph$ at $R=0.4$~m is also provided for a separate shot in vacuum. (c) Time series of $\partial\bz/\partial t$ (representative of $\jph$) evaluated along a 90$\degrees$ arc in the $\phi$ direction.}
	\label{f: erec}
\end{figure}

With the knowledge of the separator trajectory $R_X(t)$, the absolute rate of magnetic reconnection can be inferred directly from the profile of $\Psi(t,R)$. Assuming axisymmetry (which will be justified below), the amount of magnetic flux below the separator along its trajectory is $\Psi_X=\Psi(t,R_X(t))$. As shown by the red line in figure~\ref{f: erec}(a), $\Psi_X$ decreases in time corresponding to the inductive loop voltage along the separator $- d \Psi(t,R_X(t))/dt$, which represents the rate of reconnection of the full system. Then, by Faraday's law, the reconnection rate per unit length toroidally along the X-line is given by the inductive electric field
\begin{equation}
\erec(t) = -\frac{1}{2\pi R_X(t)} \frac{d \Psi(t,R_X(t))}{dt} \, .
\label{eq: erec}
\end{equation}

The inferred profile of $\erec(t)$ is shown by the red line in  figure~\ref{f: erec}(b) and is observed to be near constant in time. This is in contrast to the behavior of the inductive electric field observed in the lab-frame at $R=0.4$~m (magenta line) during a plasma discharge which is strongly modified from the vacuum electric field observed with the same drive (black line). As is evident by the magenta line, the reconnection drive from 17~{\mus}$<t<$19~{\mus} is fully shielded by the plasma current channel building close to the coils as they are first energized. Then, once the reconnection geometry fully forms and travels past $R=0.4$~m, the magenta line follows a characteristic wave pattern about the black vacuum-line corresponding to the perturbation in the inductive electric field from the steadily moving reconnection current layer. At $t\simeq31$~{\mus}, the current layer peak has reached $R=0.4$~m, and the magenta and red traces coincide. Therefore, at the X-line, the reconnection electric field $\erec$ is identical to the inductive electric field observed in the lab-frame, which is expected here since $B_{Z}=0$.


The analysis above relies on an assumption of toroidal symmetry. This symmetry has been experimentally verified using a specially built curved $\bdot$ probe which is inserted into the plasma at $R=0.4$~m and near the $Z$ location of the X-line. The probe array then measures $\partial B_Z/\partial t$ with magnetic pickup loops along a 90$\degrees$ arc in the $\phi$ direction. The data in figure~\ref{f: erec}(c) shows how the reconnection layer reaches the curved probe simultaneously (at $t\simeq32$~{\mus}) along its full toroidal arc. Some structure and variation observed in $\partial B_Z/\partial t$ in the vicinity of the reconnection layer may be associated with a lower-hybrid drift instability and is the subject of continuing research on TREX. Nevertheless, the simultaneous observation of the peak current along the full probe is direct evidence that the reconnection layer propagates inwards with strong toroidal symmetry. This data therefore supports the assumption of toroidal symmetry in evaluating $\Psi(t,R_{X}(t))$, from  which the toroidal inductive electric field, $\erec$ is inferred. Note, however, that additional electrostatic electric fields may still be present in the toroidal direction, but their integral along the full X-line will vanish as electrostatic fields always have $- \oint \nabla \Phi \cdot d{\bf l} =0$. 


\section{Shock Front and Magnetic Flux Pileup} \label{sec: shock}

\begin{figure}
	\centering
 \includegraphics[width=0.75\textwidth]{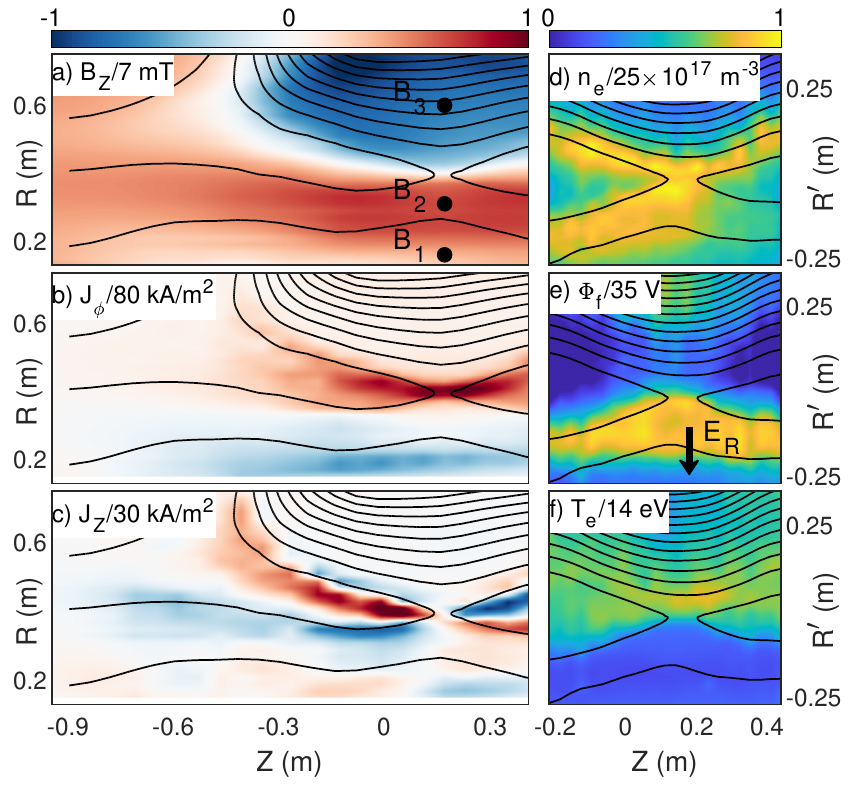}
	\caption[Reconnection profiles with prominent shock front]{Profiles of a reconnection discharge in $\hgas$ with $\bh=5$ mT and $\vdr=5$ kV. (a-c) The reconnecting magnetic field $\bz$, toroidal current density $\jph$, and poloidal current density $\jz$ are reconstructed from a radial scan with the hook probe array. Contours of constant $\Psi$ are overlaid in black representing the poloidal magnetic field lines. (d-f) The electron density $n_e$, plasma floating potential $\Vf$, and electron temperature $T_e$ measured by scanning the electrostatic probe along $Z$ at $R=0.4$~m. Here, $R'=t' \vlay$ indicates the radial profile inferred from the temporal probe signal centered around $R=0.4$~m.}
	\label{f: rec}
\end{figure}

The TREX experiment operates in a low collisional regime where Hall physics becomes important \citep{Olson2016}. An example event is presented in figure~\ref{f: rec} in $\hgas$ with $\vdr=5$~kV and $\bh=5$~mT. The data in figure~\ref{f: rec}(a-c), taken from figure~\ref{f: hook} at $\Delta t=4$~{\mus}, includes the reconnecting magnetic field component $\bz$ as well as profiles of the toroidal current $J_{\phi}$ and in-plane Hall current $J_Z$. The steady motion of the reconnection layer demonstrated in figure~\ref{f: hook} and the nearly constant layer speed shown in figure~\ref{f: psi} facilitates the ``jogging'' method applied to obtain the profiles of electron density $n_e$, plasma floating potential $\Vf$, and electron temperature $T_e$ in figure~\ref{f: rec}(d-f) measured by the $T_e$ probe at fixed $R=0.4$~m and varying $Z$ over multiple discharges. The time traces at each $Z$ position are converted to a radial chord of measurements with
\begin{equation}
	R^{\prime}=R-R_0=(t-t_0) \vlay  \, ,
\end{equation}
where $t_0$ is the time the current layer passes $R_0=0.4$~m. The density profile is asymmetric with larger densities at low $R$. In the vicinity of the reconnection region the floating potential in figure~\ref{f: rec}(e) has a structure similar to that observed by the Cluster mission during reconnection in the Earth's magnetotail \citep{Wygant2005}. In addition, a sharp jump in the potential structure, $\Delta \Vs\simeq15$~V, is observed at $R'\simeq -0.2$~m providing evidence of a collisionless shock normal electric field $E_R$. 

\begin{figure}
	\centering
	\includegraphics[width=\textwidth]{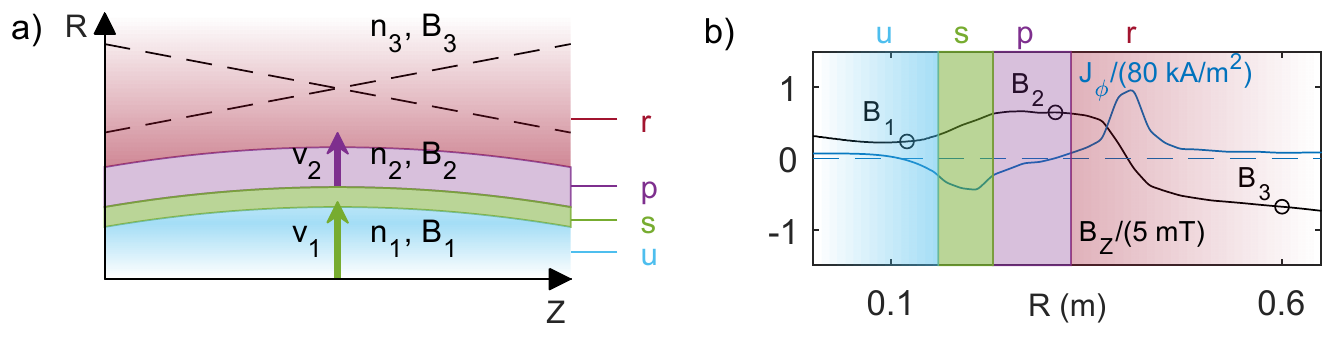}
	\caption[The shock interface on TREX]{(a) A cartoon depiction of different regions within a TREX discharge. As the current layer is driven into the background plasma, a shock interface (green) separates a region of pileup (purple) from the far upstream (blue), preceding the reconnection layer (red). (b) Radial profiles of $\bz$ and $\jph$ along a cut through the X-line ($Z=0.17$~m) in figure~\ref{f: rec}.}
	\label{f: shock}
\end{figure}

The far upstream plasma (low $R$) acts as the leading inflow, and its speed, in the frame of the reconnection layer, typically exceeds the local Alfv\'en speed. This upstream inflow must therefore be throttled in order for reconnection to take place at the appropriate rate. Thus, a region of magnetic pileup is observed to interface between the reconnection and upstream regions. Figure~\ref{f: shock}(a) shows a diagram of this configuration in which the upstream region, denoted by the color blue, is separated from the red reconnection layer by a shock in green and a region of pileup in purple. The differences are clear in figure \ref{f: shock}(b) where an increase from the upstream field corresponds with a negative spike in toroidal current, driven by the electron $E\times B$-drift with $E_R$ and $B_{Z}$. The surface current density across the shock can then be estimated as $K=e n_e\Delta\Vs/B
\simeq 1.6~\mathrm{kA/m}$,
which is in agreement with the observed magnetic pileup of 
$\Delta B=\mu_0 K\simeq 2~\mathrm{~mT}$
across the shock. The radial electric field $E_R$ is also responsible for reducing the speed of the incoming ions traveling across the shock.

As a consequence of the high Lundquist number regime ($S>10^3$) in which TREX operates, the negative toroidal current is not expected to be an inductive response to the inward-moving reconnection current sheet. Outside the diffusion region, the pressure tensor and electron inertia terms in Ohm's law can be neglected such that ${\bf E} + {\bf v_e} \cross {\bf B} \simeq \eta {\bf J}$. Using the experimental parameters within the shock we find $\eta  J_{\phi}  \simeq 5\cdot 10^{-5}~\Omega \mathrm{m} \times   30~\mathrm{kA/m^2} = 1.5 ~\mathrm{V/m}$,
which is much smaller than the typical inductive electric field $|E_{\phi}| \gtrsim 50$~V/m. Therefore, in the shock layer, ${\bf E} \simeq -{\bf v_e} \cross {\bf B}$ and the role of $E_\phi$ is to drive radial flows of the electron fluid such that this fluid strictly follows the radial motion of the magnetic field lines.

\begin{figure}
	\centering
	\includegraphics[width=0.65\textwidth]{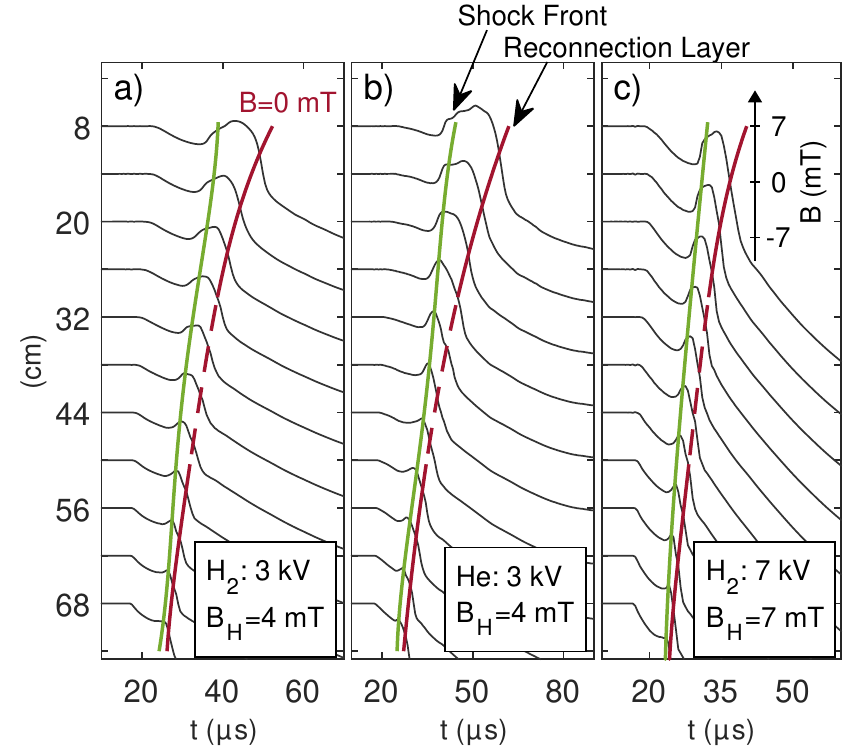}
	\caption[How scan parameters affect the reconnection dynamics]{(a-c) Stack plots of magnetic field versus time from the speed probe for three different cases. The scales are normalized such that 1~mT$=$1~cm and offset by the $R$ location of each probe. The changing time axes are indicative of the difference in timescales for these discharges. The red line follows the $B=0$~mT contour while the green line indicates the shock front leading the current layer.}
	\label{f: traces}
\end{figure}

The profiles in figure~\ref{f: shock} provide a qualitative picture of the TREX configurations whereas the exact details, how fast each layer moves or the amount of pileup, depend both on the reconnection dynamics as well as the imposed experimental settings. The gas species, number of plasma guns, Helmholtz field, and drive voltage all control certain aspects of a reconnection event. Figure~\ref{f: traces} highlights how the magnetic field evolves throughout a shot for three different cases with varying gas, $\bh$, and $\vdr$. Each individual probe trace from the speed probe is normalized to 1~mT and offset by its respective radial location, mapping out the magnetic field as the layer moves past each probe toward the central axis. The solid red line indicates the trajectory of the layer where $\bz=0$~mT. The red dashed segment marks the region of interest corresponding to $R=0.4$~m, the radial location of other probes, and where the following analysis is completed. Compared to figure~\ref{f: traces}(a) with $\vlay=33$~km/s, the layer in figure~\ref{f: traces}(b) only moves at 22~km/s, indicative of the heavier ion species. Additionally, both $\vlay$ and $\brec$ increase with $\vdr$ when comparing figure~\ref{f: traces}(a,c). In each case, a shock front, indicated by the green line, propagates at a speed $\vpile>\vlay$ ahead of the current layer as it is driven into the relatively stationary bulk plasma. The shock compresses the upstream field to match the allowable reconnection rate, setting how quickly magnetic flux is transferred through the diffusion region, and thereby resulting in the observed $\brec$ and $\vlay$.

Figure \ref{f: traces} presents a subset of the larger overall scan performed on TREX consisting of 882 total shots. All possible combinations of gas ($\hgas$, D, and He) and number of plasma guns (2 and 6 guns) with sets of Helmholtz field (2-7~mT) and drive voltage (1-8~kV) were explored, encompassing 90 different configurations of $\sim$10 shots each. Sets of $\bh$ and $\vdr$ were chosen to roughly balance each other in order to create reconnection layers that are only slightly asymmetric. Extreme cases of $\vdr$ relative to $\bh$ (or the inverse) are not considered here.

In previous flux pileup studies \citep[see][]{Oieroset2019, Dorelli2003a}, the areas where the magnetic field is increasing are considered a part of the pileup region. Similarly, we consider the observed shock formation as an integral part of the pileup dynamics. As will be detailed in \S\ref{sec: rate}, while the drive imposes the absolute reconnection rate, it is the pileup magnetic field that largely controls the normalized rate of reconnection
such that a stronger pileup magnetic field yields a lower normalized rate.
Furthermore, the reconnection rate adheres to an intrinsic scaling that is a function of the normalized system size, $L/d_i$, and it is this scaling that sets the pileup field magnitude. The shock formation described in the present section is then required to facilitate force balance between the far upstream and pileup regions in a way that is consistent with the Rankine-Hugoniot jump conditions for shocked flows \citep{Rankine1870, Hugoniot1887, Hugoniot1889, Kennel1985}. 

To properly account for the flowing mass into and out of the shock interface, the ram pressure in addition to magnetic and plasma pressure must be included in the total pressure,
\begin{equation}
	P_\alpha = m_i n_\alpha v_\alpha^2 + \frac{B_\alpha^2}{2\mu_0} + n_\alpha T_{\alpha} \, ,
	\label{eq: pressure}
\end{equation}
where the subscript $\alpha$ corresponds to individual regions indicated in the simplified drawing in figure~\ref{f: shock}(a).
Flux and particle continuity provide further constraints on the allowable parameters such that
\begin{equation}
\label{eq: nvb}
\frac{n_1}{B_1} = \frac{n_2}{B_2} \,\, , \quad
\frac{n_1}{v_2} = \frac{n_2}{v_1} \,\, ,
\end{equation}
where subscripts 1 and 2 correspond to the upstream and pileup regions, respectively. Using \ref{eq: pressure} and \ref{eq: nvb}, the total pressures for both regions across the shock interface are then
\begin{equation}
\label{eq: pressures}
P_1  = m_i n_1 v_1^2 + \frac{B_1^2}{2\mu_0} + n_1 T_{1} \,\,,\quad
P_2  = m_i n_1 v_1^2 \frac{B_1}{B_2} + \frac{B_2^2}{2\mu_0} + n_1 T_{2} \frac{B_2}{B_1}\,\, ,
\end{equation}
where $T=T_e+T_i \simeq T_e$ throughout the experiment. The velocities in \ref{eq: pressures} are described in the frame of the shock layer, but $v_1$ can simply be taken as $\vpile$, the shock velocity measured in the lab frame. Typical values for $\vpile$ range from $1.1\ms$ to $1.5\ms$, where $\ms$ is the magnetosonic Mach number. As seen in figure~8 of \citet{Kennel1985}, for the observed plasma beta of $\beta \sim 0.1$ on TREX, the critical Mach number at which all ions will be reflected from an incoming perpendicular shock is~$\sim$2.7, well above the observed range of $\vpile$ for these experiments. Additionally, the electron inertial length $d_e\sim1$~cm at the shock front is larger than the characteristic magnetic Reynolds length scale, $L_m=\eta/(\perme \vpile)\sim0.1$~cm. 
Therefore, the observed shocks tend to steepen to the size of a few $d_e$ wide, consistent with a sub-critical dispersive (rather than resistive) shock \citep{Kennel1985}. 

\begin{figure}
	\centering
	\includegraphics[width=0.82\textwidth]{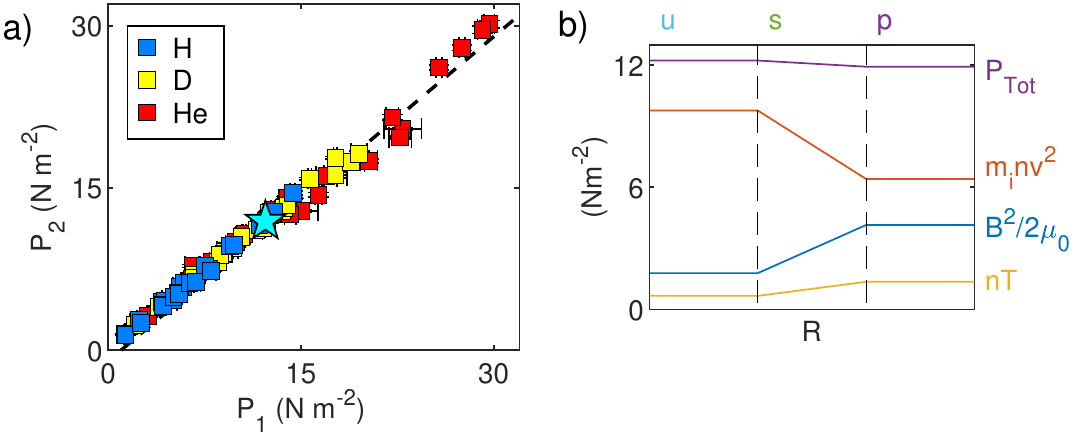}
	\caption[Pressure balance across the shock front]{(a) A comparison of the total pressures $P_1$ and $P_2$ for all configurations in the dataset. In many cases, the standard deviation error falls below the marker size. (b) Depiction of the change in total pressure along with ram, magnetic, and plasma pressures for the case indicated by the blue star (the event in figure~\ref{f: rec}) going from the upstream (u) to the pileup (p) regions across the shock interface (s).}
	\label{f: shock 2}
\end{figure}

The upstream and pileup pressures are easily computed from measurements and are plotted in figure~\ref{f: shock 2}(a) where each data point corresponds to the average of all shots for a single configuration in the parameter scan. The dashed line of slope 1 shows good agreement between the total pressure on either side of the shock front and that force balance is satisfied across the interface regardless of the externally imposed conditions. Pressure profiles for the configuration in figure~\ref{f: rec} are given in figure~\ref{f: shock 2}(b) which shows a discrete drop in ram pressure across the shock front that is largely balanced by a rise in magnetic pressure such that the total pressure remains nearly constant. 

For this set of experiments, shocks are observed to form in all Hydrogen and Deuterium cases and most of the Helium cases. Deriving a strict drive-threshold for the shock formation is difficult with the given dataset as the background magnetic field and drive voltage are not fully independent. Rather, $\bh$ is adjusted within a range that decreases with decreasing reconnection drive. Furthermore, after triggering the reconnection drive but before the reconnection and shock layers have fully formed, the background field declines, which is most pronounced at higher $\vdr$. The upstream Alfv\'en speed of the shock is then indirectly related to the reconnection drive. In other experiments not presented here, shock formation is not found to occur for $\vdr<300$~V in Hydrogen.

The TREX configuration is perhaps unusual in the sense that reconnection is only driven from one side of the reconnection layer. Because of the strong magnetic fields associated with the drive, on the side where the drive is applied the Alfv\'en speed becomes large.  Meanwhile, for the undriven side, the Alfv\'en speed is low and allows for supersonic flows and shock formation not seen in symmetrically driven experiments. While shock fronts on either side of a reconnection layer could easily be envisioned in space and astrophysical situations where supersonic plasma winds collide, such configurations are not easily obtained in laboratory experiments relying on a magnetic drive. It is possible that they can be obtained in laser-driven reconnection experiments, but here the plasma beta is often so large that the magnetic field plays a limited role in setting the overall plasma dynamics  \citep{Fox2011}. 

\section{Reconnection Rate} \label{sec: rate}

In conjunction with the pileup front, the reconnection region also freely develops given the experimantal conditions, typically exhibiting both magnetic and density asymmetries across the layer. To account for this, flux and particle continuity into the layer can be used to derive an expected scaling for asymmetric reconnection \citep{cassak2007} dependent on both inflow magnetic field and density values $B_2,~B_3,~n_2,~\textrm{and}~n_3$. 
The expected reconnection rate is then written as 
\begin{equation}\label{eg:1}
\erec=\alpha\vah\bred \, ,
\end{equation}
where $\alpha$ is a factor dependent on the specific geometry of the layer. This is similar to the symmetric reconnection rate but modified by a reduced magnetic field
\begin{equation}
	\bred=\frac{2B_2B_3}{B_2+B_3}
\end{equation}
and a hybrid Alfv\'en speed,
\begin{equation}
\vah=\left( \frac{1}{{\mu}_0 m_i}\frac{B_2B_3(B_2+B_3)}{n_3B_2+n_2B_3} \right)^{1/2} \, ,
\end{equation}
where $m_i$ is the ion mass. Typically, $B_2<B_3$ while $n_3<n_2$. 	As indicated in figures~\ref{f: rec}(a) and \ref{f: shock}(b), values for $\vah$ and $\bred$ are determined by choosing measurements $\sim1d_i$ away from either side of the layer, consistent with procedures used in theory \citep{Shay2001}. The location of the flux pileup values naturally sits~~$\gtrsim0.6d_i$ away from the reconnection layer.

\begin{figure}
	\centering
	\includegraphics[width=\textwidth]{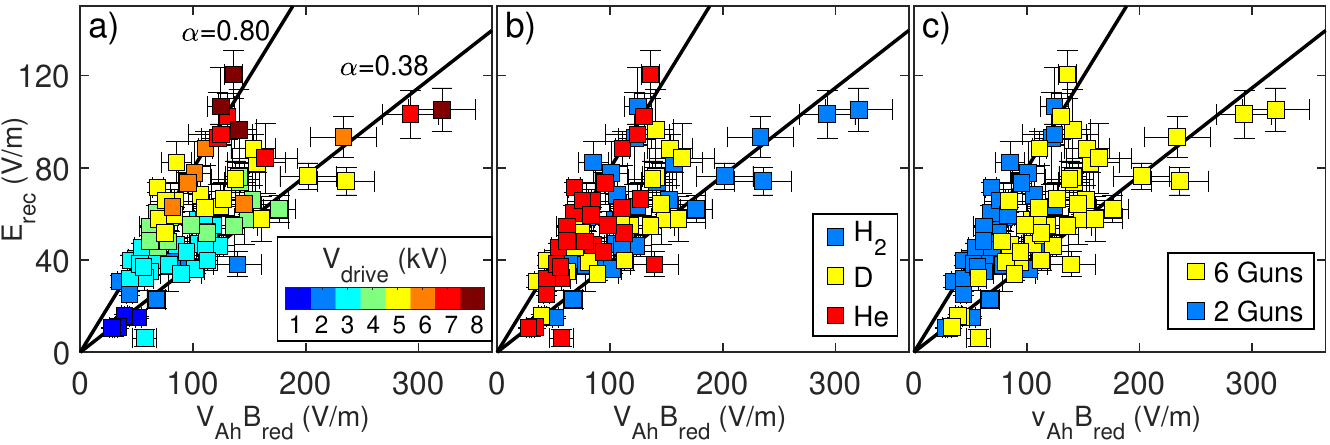}
	\caption[Reconnection rate scaling on TREX]{(a) Scaling of the reconnection rate with respect to $\vdr$. Each data point represents the average of all shots for a single configuration with errorbars indicating the weighted variance for each group. The dataset ranges in slope between $\alpha\approx0.38\text{--}0.8$, where $\alpha=\erec/\vah\bred$ represents the normalized reconnection rate. The same dataset as in (a) is shown with respect to ion species (b) and number of plasma guns (c). }	
	\label{f: rate}
\end{figure}

As described in \S\ref{sec: erec}, the measured reconnection electric field $\erec$ is readily computed by finding the change in magnetic flux along the X-line trajectory (\ref{eq: erec}). Carrying out the above analysis for all shots in the scan provides an overall scaling of $\erec$ to $\vah\bred$ for the TREX experiment, shown in figure~\ref{f: rate} with data points colored by different parameters. The slope of the data represents the normalized reconnection rate
\begin{equation}
	\alpha=\frac{\erec}{\vah\bred}
\end{equation}
with the two solid lines showing that the data falls between a rate of $\alpha\approx0.38\text{--}0.80$. These values for $\alpha$ are larger than the rate $\alpha\sim 0.1$ typically associated with fast reconnection. From figure~\ref{f: rate}(a), it is clear that the absolute electric field shows significant dependency on $\vdr$ where $\erec$ increases roughly proportionally as $\vdr$ is increased from 1 kV to 8 kV. Figures~\ref{f: rate}(b-c) show the same results with respect to the gas and number of plasma guns used, respectively, as the ion species and density are included in the $\vah$ term. Here the number of plasma guns acts as a proxy for density with a factor of two difference in density between similar configurations running six versus two guns. While there is a distinct bifurcation in the normalized rate between the six and two gun cases, there is still some variation in the results seen in figure~\ref{f: rate}(c). Meanwhile, figure~\ref{f: rate}(b) shows spread between the two bounds in the rate for all gases used in the scan.

\begin{figure}
	\centering
	\includegraphics[width=0.85\textwidth]{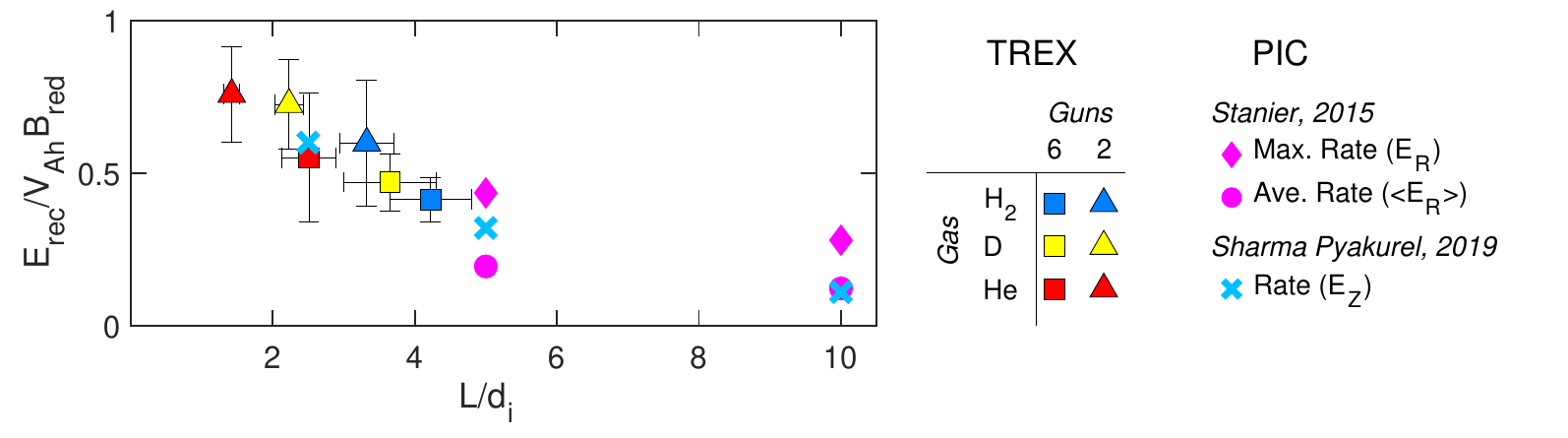}
	\caption[Reconnection rate versus system size]{The reconnection rate $\alpha$ as a function of the normalized system size $L/d_i$, where $L\simeq0.8$ is the half-length of the current layer. Each data point represents all discharges with similar ion species and number of plasma guns. Included in pink and light blue are measured reconnection rates from recent PIC simulations of colliding flux ropes from \citet{Stanier2015a} and turbulent magnetic bubbles from \citet{SharmaPyakurel2019}, respectively.}
	\label{f: size}
\end{figure}

The data on the reconnection rate is, perhaps, more appropriately considered in the context of the system size compared to the relevant ion kinetic length scale, or $d_i$ for the case of anti-parallel reconnection. Another representation of the results is thus provided in figure~\ref{f: size}, displaying $\alpha =\erec/(\vah\bred)$ as a function of the relative system size $L/d_i$, where $L\simeq0.8$~m is the half-length of the current layer imposed by the TREX drive coils. The ion masses and observed densities correspond to a range of roughly $1\lesssim L/d_i \lesssim 5$ where the upper and lower bounds are associated with the highest density Hydrogen and lowest density Helium cases, respectively. Figure~\ref{f: size} shows that the reconnection rate increases as the relative system size decreases. While the reconnection rate remains large for this set of experiments, it does not conflict with the expected rate for fast reconnection of $\alpha \sim 0.1$. Considering the the relatively small system size ($L<5d_i$), these results are most closely related to studies of island coalescence (flux rope mergers) or turbulent magnetic reconnection. Results from fully kinetic PIC simulations with magnetic islands down to $L=2.5d_i$ are also included from \citet{SharmaPyakurel2019} and \citet{Stanier2015a} in figure~\ref{f: size}, showing the TREX data maintains the trend of increasing $\alpha$ for smaller system size. A rate approaching $0.1$ is expected for TREX experiments which reach a larger system size of $L/d_i\gtrsim10$ by increasing the imposed length $L$ and decreasing the realized $d_i$ in the experiment.

\section{Discussion} \label{sec: disc}

The results presented thus provide experimental confirmation of previous numerical studies on island coalescence, as well as flux pileup. Simulations of laser produced plasma bubbles have shown that the reconnection rate only matches the expected rate when taking into consideration flux pileup \citep{Fox2011} with system size of $L\approx20d_i$, while corresponding laboratory experiments see rates of $\alpha \sim 1$ \citep{Fiksel2014} attributed to the transient behavior of the systems. As a main difference, the observed reconnection rate on TREX is less transient in nature and for smaller system size. As is evidenced in figure~\ref{f: rec}, the reconnection geometry is steadily moving as the lowest field line travels from the inflow to X-line, thus approximating several Alfv\'enic crossing times in the timeframe shown. This, in conjunction with the nearly constant $\erec$ in figure~\ref{f: erec}, provides evidence of ample time for the reconnection process on TREX to reach a near steady state.

To emphasize  the main experimental findings we summarize a few key observations. The TREX configuration is implemented with a strong reconnection drive, mainly motivated by the desire to study reconnection in a  high Lundquist number regime, $S >10^3$. As shown in figure~\ref{f: rate}(a), the strength of this  drive roughly sets the absolute rate of reconnection. With the strong drive, the imposed absolute reconnection rate is so large that without the enhancement of the magnetic field in the low-$R$ reconnection inflow region, the normalized reconnection rate would often be larger than 10. Such a large normalized rate is naturally unphysical and would  not be in agreement with previous studies of reconnection. Instead, the magnetic field piles up, which yields larger values of $B$, providing the enhanced tension  of the upstream magnetic field required for reconnection to process the inertia of the ion fluid as it is driven into the exhaust. Thus, the upstream value of $B$ increases until the corresponding Alfv\'en speed is sufficiently large that the ratio $\alpha=\erec/(\vah\bred)$ is reduced to be consistent with the intrinsic normalized rate of reconnection, which, as evident by the results in figure~\ref{f: size}, is regulated by the normalized system size. For this reason, the TREX results are not in conflict with studies of island coalescence which show reduced pileup for small system size $L\sim5d_i$ \citep{Karimabadi2011}. This reduction is likely due to the intrinsic reconnection drive (the attraction between current channels) which decreases with system size, whereas the drive in TREX is always strong.

The enhanced pressure of the plasma upstream of the reconnection naturally needs to be in force balance with the plasma further upstream in the reconnection inflow. Given the relatively low Alfv\'en speed of this far upstream plasma, this force balance is achieved through the formation of the shock layer. Between the shock and the reconnection layer, the Alfv\'en speed is large such that the two structures can ``communicate'' and develop in a way that 1) force balance of the overall system is maintained and is consistent with the shock jump-conditions in figure~\ref{f: shock 2} and 2) the value of the piled-up magnetic field is just right that the normalized reconnection rate falls on the curve of figure~\ref{f: size}.  

The reconnection scenario implemented in TREX is quite different from other reconnection experiments where reconnection is driven more symmetrically from either the exhaust sides or the inflow sides \citep{Yamada1999}. Nevertheless, a range of results on reconnection are observed to be robust and not dependent on the particular scenario by which reconnection is driven. As such, it is a notable result that the normalized rate of reconnection in the asymmetrically driven TREX configuration provides a scaling law with system size fully consistent with the results from nominally very different numerical scenarios starting from the idealized Harris-sheet configuration \citep{SharmaPyakurel2019} or coalescing islands \citep{Stanier2015a}. The increasing normalized rate at smaller system size may be indicative of a transition to ``electron only reconnection'' \citep{Phan2018} where the ions do not strongly couple to the exhaust, permitting a larger electron reconnection outflow speed. In fact, at the scale of the electron diffusion region, the electron exhaust velocity approaches the electron Alfv\'en speed \citep{Drake2008}. As the system size increases, the ions become increasingly more coupled, thus reducing the normalized reconnection rate and approaching those expected for MHD systems \citep{Liu2017}. Further comparisons to theoretical predictions of ``electron only reconnection'' are the subject of future experimental campaigns on TREX. 
\\

We gratefully acknowledge DOE funds DE-SC0019153, DE-SC0013032, and DE-SC0010463 and NASA fund 80NSSC18K1231 for support of the TREX experiment. In addition, this work is supported through the WiPPL User Facility under DOE fund DE-SC0018266.

\bibliographystyle{jpp}
\bibliography{references.bib}

\end{document}